# Title: Bright excitons with negative-mass electrons

**Authors:** Kai-Qiang Lin[1]*, Chin Shen Ong[2,3], Sebastian Bange[1], Paulo E. Faria Junior[1], Bo Peng[4], Jonas D. Ziegler[1], Jonas Zipfel[1], Christian Bäuml[1], Nicola Paradiso[1], Kenji Watanabe[5], Takashi Taniguchi[5], Christoph Strunk[1], Bartomeu Monserrat[4,6], Jaroslav Fabian[1], Alexey Chernikov[1], Diana Y. Qiu[2,3,7], Steven G. Louie[2,3], John M. Lupton[1]*

**Affiliations:**

[1]Department of Physics, University of Regensburg, Regensburg, Germany.

[2]Department of Physics, University of California at Berkeley, CA, USA.

[3]Materials Sciences Division, Lawrence Berkeley National Laboratory, Berkeley, CA, USA.

[4]Cavendish Laboratory, University of Cambridge, Cambridge, UK.

[5]National Institute for Materials Science, Tsukuba, Ibaraki, Japan.

[6]Department of Materials Science and Metallurgy, University of Cambridge, Cambridge, UK.

[7]Department of Mechanical Engineering and Materials Science, Yale University, CT, USA.

*Correspondence to: kaiqiang.lin@ur.de, john.lupton@ur.de

**Abstract:**

Bound electron-hole excitonic states are generally not expected to form with charges of negative effective mass. We identify such excitons in a single layer of the semiconductor $WSe_2$, where they give rise to narrow-band upconverted photoluminescence in the UV, at an energy of 1.66 eV above the first band-edge excitonic transition. Negative band curvature and strong electron-phonon coupling result in a cascaded phonon progression with equidistant peaks in the photoluminescence spectrum, resolvable to ninth order. *Ab initio GW*-BSE calculations with full electron-hole correlations unmask and explain the admixture of upper conduction-band states to this complex many-body excitation: an optically bright, bound exciton in resonance with the semiconductor continuum. This exciton is responsible for atomic-like quantum-interference phenomena such as electromagnetically induced transparency. Since band curvature can be tuned by pressure or strain, synthesis of exotic quasiparticles such as flat-band excitons with infinite reduced mass becomes feasible.

**One Sentence Summary:** Unconventional high-lying bound excitonic states comprising negative-mass electrons and positive-mass holes are found in monolayer $WSe_2$, which shows radiative recombination in the UV at an energy of almost twice the bandgap.

**Main Text:**

Mass is perhaps the most tangible and the most elusive concept of physics. For stable matter, mass is strictly positive. In quantum mechanics, however, interacting electrons in a medium can behave like free particles with an effective mass related to the curvature of their energy-momentum dispersion relation. In periodic potentials, such as that of the crystal lattice of a semiconductor, the effective mass can become negative in certain parts of the bands, implying counterintuitive effects such as a tendency for opposite charges to accelerate apart rather than to attract via the Coulomb interaction. At sufficiently low temperatures, an electron excited to the conduction band and a hole in the valence band of a semiconductor can interact to form a stable correlated state – an exciton. Typically, excitons with stable bound states are made up of a positive-mass electron in the lowest-energy conduction bands and a positive-mass hole in the highest valence bands. In monolayer $WSe_2$, such excitons can exhibit large optical transition dipoles and thus interact strongly with incident radiation (*1, 2*). When driven by an intense laser pulse, discrete band-edge excitons in monolayer $WSe_2$ have been found to exhibit signatures of quantum interference in their transition pathways (*3*). In addition, efficient Auger-like upconversion has been attributed to transitions to higher conduction bands (*4*). These experiments suggest that the lowest-energy, optically bright exciton couples strongly to a discrete long-lived state of approximately twice the energy of the lowest band-edge exciton. The resulting multilevel excitonic state structure of $WSe_2$ is therefore reminiscent of that of optically driven atomic systems exhibiting electromagnetically induced transparency (EIT) (*5*). This purported high-energy state appears to behave like a bound – or very strong resonant – exciton, but it is not clear how such an exciton would remain stable for timescales exceeding 100 fs as implied by the EIT phenomenon (*3, 6*), at energies reaching far into the free-particle continuum around the K-points. Such an exciton appears particularly improbable when the independent-particle interband transition picture suggests that the only available electronic states with consistent energy and momenta are of negative effective mass.

Figure 1A sketches the calculated *ab initio GW* quasiparticle band structure of monolayer $WSe_2$ around the K-points in momentum space. The magnitude of the interband transition oscillator strength coupling the spin-orbit split top valence band to the different conduction bands is coded in color. The full band structure is shown in Fig. S1A. The lowest-energy band-edge "A-exciton" is characterized by a dipole-allowed optical transition between the valence-band maximum and the upper spin-split conduction-band minimum at the K-points in the Brillouin zone. The A-exciton can recombine radiatively, emitting a photon with an energy reduced with respect to the quasiparticle bandgap by the exciton binding energy, which can be as large as a fraction of an eV for transition-metal dichalcogenide (TMDC) monolayers (*1, 2*). The energy of this transition is indicated by the lower red double-headed arrow. Signatures of excitonic quantum interference are observed in optical second-harmonic generation (SHG) under pumping by ultrashort pulses with photon energies just below the energy of this transition (*3, 6*), suggesting the involvement of a second excitonic state approximately 1.7 eV above the A-exciton (higher-energy red double-headed arrow). Within the independent-particle picture, the only electronic state close to this energy range is the lower spin-split CB+2 band in Fig. 1A. This band has negative curvature. Our calculations in Fig. 1A confirm that optical transitions from this band ($CB+2^-$) to the top valence band ($VB^+$) are allowed (*2, 7*), with approximately 4% of the oscillator strength of the band-edge transition ($CB^+$ to $VB^+$) at the K-points. While the idea of a stable exciton involving a negative-mass electron seems counterintuitive, such a complex is possible within a simple effective-mass hydrogenic model provided the hole mass $m_{\mathrm{h}}^*$ is *positive* and smaller in magnitude than the negative electron mass $m_{\mathrm{e}}^*$. The reduced mass of the exciton ($1/\mu = 1/m_{\mathrm{h}}^* + 1/m_{\mathrm{e}}^*$) is then positive (see Fig. S2). Drawing an analogy to classical orbital motion, the semiclassical motion of the electron and hole may then be thought of as orbiting around a common center which does not lie geometrically between the two particles, so that, as indicated in the inset of Fig. 1B, the electron accelerates in the same direction as the hole.

A crucial difference between excitons comprising electrons of positive and negative effective mass lies in the coupling to phonons: since the band-edge exciton forms at the energetic minimum, the exciton cannot dissipate energy further to a final state by emitting phonons – luminescence is therefore generally dominated by a "zero-phonon" line. The opposite is true for an exciton involving an electron from a band of negative curvature since it is in resonance with the finite-momentum electron-hole continuum: electrons tend to scatter into lower-energy band states by emitting phonons. Assuming sufficient radiative decay rates (*8*), a phonon progression should appear in the luminescence and the zero-phonon transition needs not be the maximum.

To selectively excite electron-hole pairs in the vicinity of the fundamental bandgap at the K-points, we use a linearly polarized narrow-band continuous-wave (CW) laser at 716 nm (1.732 eV), resonantly driving the A-exciton of monolayer $WSe_2$ encapsulated in hexagonal boron nitride (hBN) (see Fig. S3 for sample details). As sketched in the bottom-left inset of Fig. 1C, Auger-like exciton-exciton annihilation (*4, 9, 10*) can raise the electron to a higher band as momentum conservation localizes it around the K-points. The resulting upconversion photoluminescence (UPL) spectrum at 5 K (Fig. 1C) shows a narrow peak at twice the excitation energy, i.e. at 3.46 eV, limited in width by the spectral resolution of the monochromator. This feature arises from CW SHG (*11*) originating from the broken inversion symmetry of monolayer $WSe_2$. At an energy 60 meV below this peak, UPL is observed. Ten narrow peaks are resolved with a mean linewidth of 11.4 meV. SHG and UPL are discriminated by measuring the change in emission intensity copolarized with the laser as the laser polarization is rotated with respect to the $WSe_2$ crystal (*12*). The right inset of Fig. 1C shows the characteristic six-fold symmetry of the SHG polarization dependence, arising from the three-fold rotational crystal symmetry. In contrast, UPL is isotropic and appears as a circle. We label this emissive species the "high-lying exciton" HX. The linewidth of the dominant HX peak can be as narrow as 5.8 meV at low pump fluences as shown in Fig. S3C, and places a lower limit on the exciton coherence time of ~100 fs. Resolving such narrow discrete luminescence peaks at almost twice the bandgap is unexpected since excitonic linewidth generally increases with transition energy as plotted in Fig. S3; higher-lying transitions such as the B, A' and B' excitons (*13*) are typically subject to a broader range of effective non-radiative relaxation channels, reducing lifetime. The connection between HX UPL and the A-exciton transition can be established by sweeping the excitation energy over the A-exciton resonance. The top-left inset of Fig. 1C shows the PL spectrum of the fundamental band-edge exciton (red line) in comparison to the PL excitation (PLE) spectrum of the HX emission (blue dots); PL and PLE spectra are virtually identical. As discussed in Fig. S4, the UV UPL feature arises from a bright-exciton transition since it shows characteristics of in-plane dipole orientation just like the A-exciton, but unlike the spin-forbidden dark exciton with an out-of-plane dipole (*1*).

The ten peaks are spaced equally as seen in Fig. 2A, implying cascaded electron-phonon scattering at a phonon energy of 15.5±0.1 meV. Calculations of the phonon dispersion (Fig. S1B) in monolayer $WSe_2$ suggest that the progression interval corresponds to a longitudinal acoustic (LA) phonon near the M and K-points of the Brillouin zone (*14*). We propose that strong inelastic resonant electron-phonon scattering occurs between +K and −K valleys (*15, 16*), as sketched in Fig. 2B, C, and note that such intervalley scattering can occur in <100 fs through a deformation potential (*17-19*). As in previous monolayer TMDC studies, our *GW* band-structure calculations of monolayer $WSe_2$ reveal that all the conduction bands and the valence bands around the K-points are spin split because of spin-orbit coupling. The combination of mirror-plane and time-reversal symmetry dictates that spin in the out-of-plane direction is a good quantum number with opposite orientations in the +K and −K valleys (*20*). Two different phonon-scattering processes between valleys are thus conceivable to move a negative-mass electron of the high-lying exciton down in energy: one-phonon scattering (Fig. 2B), which requires a spin-flip (*21, 22*); and a double-resonance two-phonon mechanism (Fig. 2C) (*15*), which conserves spin. The two-phonon process should be favored in luminescence over the one-phonon process, although we note that the helicity of the excitation may be destroyed by the fast exchange interaction (*23, 24*) arising during the Auger-like population of the HX state. The spectrum in Fig. 1C demonstrates that even-numbered peaks are indeed more intense than odd-numbered ones, implying a higher transition probability for two-phonon processes. Repeated electron-phonon scattering not only moves the electron downwards in energy, but also away from the K-symmetric points. The spin-valley locking is then relaxed, which, for higher-lying bands, persists only over a limited region of momentum space. The alternation in peak intensity is pronounced up to peak 5. Inspection of the Brillouin zone of single-layer $WSe_2$, inset in Fig. 2A, together with the phonon progression, supports the assertion that HX PL originates from the K-symmetric points. If instead the radiative state were formed around the Γ or M point, which lack spin-valley locking, there would be no obvious reason why the phonon intensity should alternate. We note that the phonon progression can also be interpreted from the perspective of polaronic excitons. The question of whether the HX phonon progression should be rationalized in terms of excitonically bound charges, which emit phonons, or as polaronic excitons in split subbands (*16*), is mostly a matter of electron-phonon interaction strength and is left for further study.

A signature of the excitonic nature of optical transitions in TMDC monolayers is the existence of a series of excited states (*1*). Besides Auger-like excitation, it is also conceivable that two-photon absorption (TPA) can populate dark states that then undergo conversion to the radiative HX state. As sketched in Fig. 3A, for systems undergoing dipole-allowed interband transitions, TPA can address odd-parity *p*-like excitonic states, while one-photon absorption and emission probe even-parity *s*-like ones (*25, 26*). In order for the two-photon excitation to relax to the radiative HX, the TPA energy must be no lower than the HX zero-phonon transition energy, i.e. around 3.4 eV. There is no need for TPA to occur at exactly twice the energy of the A-exciton, i.e. to overlap with the one-photon resonant Auger-like mechanism in PLE. Figure 3B plots emission spectra as a function of excitation energy for a high pump power of 35 kW/cm$^2$, revealing a clear cut-off just below the transition energy of the A-exciton. TPA (orange arrow) sets in when the laser energy is slightly below the A-exciton transition (green marker). The diagonal line stems from SHG. Auger and TPA mechanisms are distinguished by their power dependencies in the inset of Fig. 3C: the intensity of two-photon PL (orange) scales quadratically with excitation power, whereas the Auger-like process (green) shows sublinear behavior because of bleaching of the A-exciton transition (*4*). Although the two PLE features are not degenerate, the PL spectra from TPA and Auger-like excitation are identical (panel C). The lower part of Fig. 3B illustrates the pump-power dependence of the PLE spectra, obtained from the integrated HX UPL intensity as a function of excitation energy. At low powers, only one PLE peak is seen, corresponding to the PL spectrum of the A-exciton and implying Auger-assisted upconversion to the HX (cf. Fig. 1C). At high powers, this feature broadens, possibly because of electronic scattering and increased screening of the excitonic electron-hole pair. The additional narrow TPA peak has a linewidth of ~12 meV, comparable to the linewidth of HX PL transitions in Fig. 3C, suggesting that TPA and PL transitions arise from excitonic states of similar nature. The complete evolution of PLE with pump power is shown in Fig. S5. To prove that the two-photon excitation and one-photon emission of HX do not simply arise from different phonon transitions of the same state, we also performed measurements on monolayer $WSe_2$ encapsulated by hBN layers of different thickness. Changing the dielectric environment of the monolayer modifies the binding energy of excitons and thus the energy separation between excitonic states (*1*), but does not affect intralayer phonons. As demonstrated in Fig. S6 and Table S1, the separation between *p*-like (two-photon transition) and *s*-like (one-photon emission) HX states varies between 27 meV and 37 meV across samples. No such change is seen in the spacing of the phonon progression, which is almost identical for all samples.

To establish the contribution of higher-lying conduction bands to the HX, we performed an *ab initio GW*-BSE calculation (*27-29*). Figure 4A shows the calculated absorbance spectrum of monolayer $WSe_2$. The calculation explains all the salient features (which are theoretically shown to be excitonic transitions) of the experimental reflectance contrast of hBN-encapsulated $WSe_2$ in Fig. 4B. The minor difference between the calculated and measured energies of excitonic transitions stems mainly from the influence of dielectric environment, i.e. the hBN encapsulation, which is not included in the calculation. As seen from Figs. 4A and B, there are many features around 3.4 eV, i.e. in the range of the experimentally observed HX. This multitude of transitions arises because both the calculated absorbance and the experimental reflectance probe the entire Brillouin zone. Resonant pumping of A-excitons, however, selectively populates excitations (including HX) around the K-points, leading to the UPL emission shown in panel C. To identify the HX in the calculation, we filter out excitons originated from transitions outside the K-valley, retaining only those formed by transitions within a range of less than 0.2 $Å^{-1}$ around the K-points (see Fig. S1A). Since pumping A-excitons predominantly populates holes at the topmost valence band, we only include this band ($VB^+$) here for clarity. Excitons with contributions from both $VB^+$ and $VB^-$ are shown in Fig. S7. The resulting absorbance spectrum restricted to the K-valleys is plotted in panel D, where we also show calculated absorbance spectra when the transitions include only a subset of the conduction bands: only CB (red), only CB+1 (yellow), or only CB+2 (blue). From this analysis, it is clear that the high-lying feature, HX, comes primarily from transitions involving CB+2, as the calculation restricted to only CB+2 closely reproduces the feature obtained when all CBs are included. Finally, the contributions of individual conduction bands to the excitons in the K-valley are shown in panel E, where the area of each disk is proportional to the integrated oscillator strengths (see SI). The HX peak around 3.35 eV stands out in the calculation, explaining the experimental data, with a dominant contribution from the conduction band $CB+2^-$. The lower conduction bands have a much smaller contribution to the HX peak due to the relatively large energy separation between higher (CB+2) and lower conduction bands (CB and CB+1) in the K-valleys. As seen in the band structure in Fig. 1, $CB+2^-$ corresponds to an electron of negative mass. The inset in Fig. 4D shows the envelope function of the HX wavefunction in momentum space. Its comparison with the A-exciton is shown in Fig. S8. Figure S8D projects the HX exciton envelope function onto the quasiparticle band structure and shows that the HX is localized around the K-points, where the effective-mass approximation is valid. The HX has an exciton radius (*1*) of 1.2 nm, which is smaller than the 1.5 nm radius of the 1*s* A-exciton, consistent with a heavier reduced exciton mass arising from the negative-mass electron.

Since the electronic structure of many semiconducting monolayer TMDCs is qualitatively comparable (*2, 7*), excitons with negative-mass electrons can be expected to be a generic feature of TMDCs. The band curvature of electrons and holes in two-dimensional semiconductors can be tuned by strain, dielectric environment, pressure, or band hybridization in (twisted) multilayers, offering opportunities to explore the underlying physics of excitons with negative-mass electrons as well as their coupling to conventional excitons with positive-mass electrons and holes. Heavy-boson phenomena may be expected to emerge when the moduli of effective electron (negative) and hole (positive) masses are tuned to be equal, giving rise to flat-band excitons with infinite reduced mass. Since the A-exciton can be directly promoted to a higher-energy exciton by optical pumping, single-layer $WSe_2$ behaves as an excitonic three-level system resembling atoms (*3*). This characteristic will enable the design of novel semiconductor-based optical parametric amplifiers, optical gain elements which do not require inversion, and ultrafast optical switches exploiting EIT (*5*).

**Acknowledgments:** The authors thank R. Huber, M. M. Glazov, and P. Merkl for helpful discussions, and S. Krug for technical support. **Funding:** Financial support is gratefully acknowledged from the Deutsche Forschungsgemeinschaft (DFG, German Research Foundation) – Project-ID 314695032 – SFB 1277 projects B03 and B05, and the Emmy-Noether Grant CH 1672/1-1; P.E.F.J. was supported by a Capes-Humboldt Research Fellowship of the Alexander von Humboldt Foundation (Grant No. 99999.000420/2016-06). The work at the Lawrence Berkeley National Lab was supported by the van der Waals Heterostructures Program through the Office of Science, Office of Basic Energy Sciences, U.S. Department of Energy under Contract No. DE-AC02-05CH11231, which provided the formulation of the theory and theoretical analyses; and by the National Science Foundation under Grant No. DMR-1926004, which provided for the GW and GW-BSE calculations; advanced software and codes were provided by the Center for Computational Study of Excited-State Phenomena in Energy Materials (C2SEPEM) at the Lawrence Berkeley National Laboratory, which is funded by the U.S. Department of Energy, Office of Science, Basic Energy Sciences, Materials Sciences and Engineering Division under Contract No. DE-AC02-05CH11231, as part of the Computational Materials Sciences Program. B.P. and B.M. were supported by the Gianna Angelopoulos Programme for Science, Technology, and Innovation and the Winton Programme for the Physics of Sustainability. Growth of hexagonal boron nitride crystals was supported by the Elemental Strategy Initiative conducted by the MEXT, Japan and the CREST (JPMJCR15F3), JST.

**Author contributions:** K.-Q.L. conceived the project and carried out the measurements with the support of S.B.; J.D.Z., K.-Q.L., J.Z., C.B., N.P., C.S. and A.C. contributed to the fabrication of samples; C.S.O., P.E.F.J., B.P., B.M., J. F., D.Y.Q., and S.G.L. contributed to the theoretical calculations; K.W. and T.T. provided hBN crystals; K.-Q.L., S.B., and J.M.L analyzed the data and wrote the paper with input from all authors. **Competing interests:** The authors declare no competing interests. **Data and materials availability:** All data is available in the main text or the supplementary materials.

**Supplementary Materials:**

Materials and Methods

Figures S1-S9

Tables S1-S2

References (*30*-*51*)

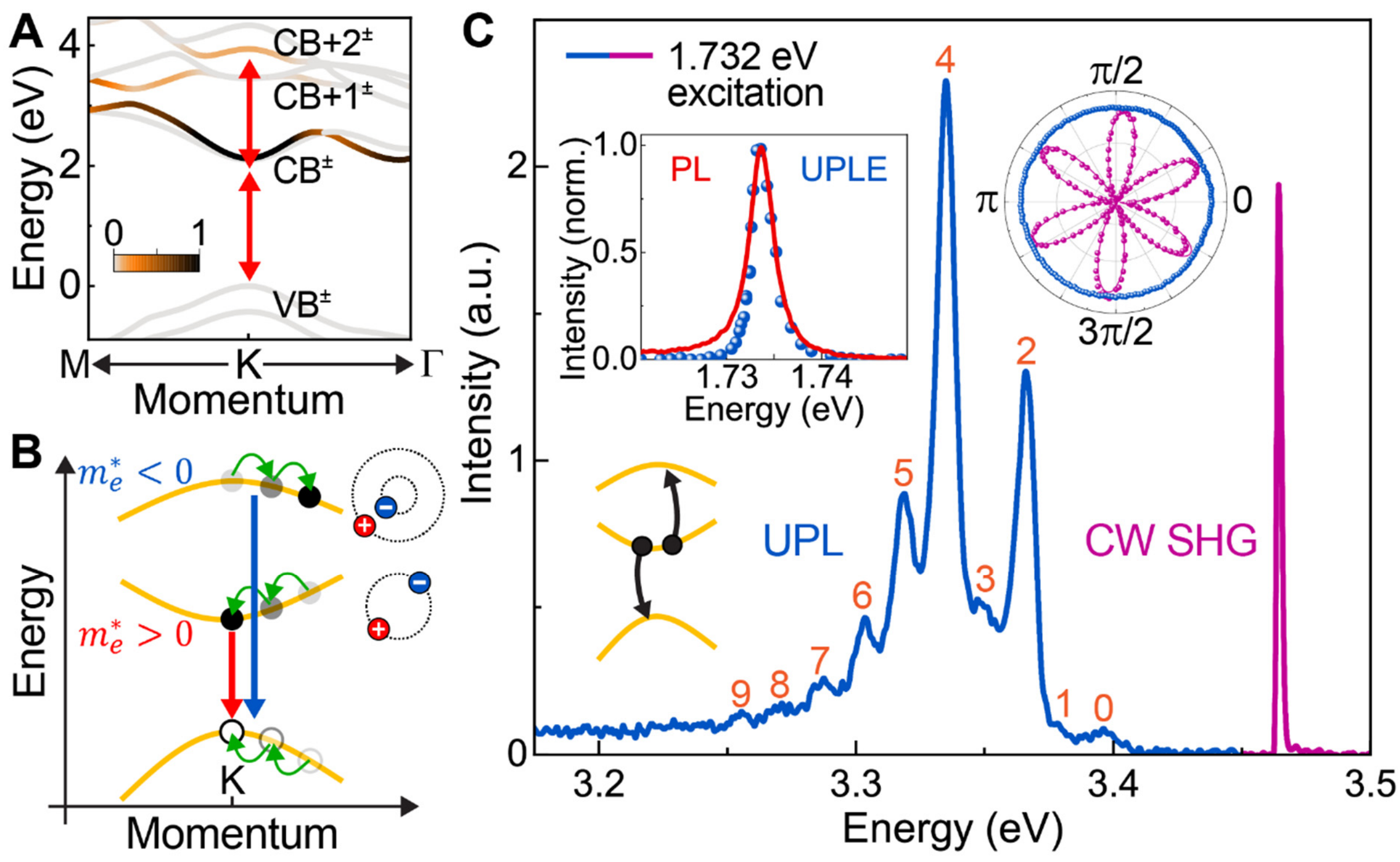

**Fig. 1. Formation of high-lying excitons (HX) in monolayer $WSe_2$, with an electron in an upper conduction band of negative curvature and a hole in the top valence band.** (A) Calculated *GW* band structure of bare monolayer $WSe_2$. Higher (lower) energy spin-split bands are labelled by + (-) superscripts. The color indicates normalized oscillator strength of electrons making a transition from the top valence band ($VB^+$) to the different conduction bands. Red double-headed arrows mark the exciton resonances invoked to explain quantum interference in second-harmonic generation (*3, 6*). (B) Schematic of optical transitions from two distinct conduction bands of opposite curvature to the valence band at the K-symmetry points. The positive curvature of the lowest conduction band (CB) implies positive effective mass of electrons close to the minimum, where an excitonic bound state of electron and hole can form – the band-edge exciton. Excited holes in the VB and excited electrons in the CB relax towards the K-points by emitting phonons, whereas the higher-band negative-mass electron relaxes away from the K-points. (C) Photoluminescence (PL) spectrum of hBN-encapsulated monolayer $WSe_2$ at 5K under narrow-band continuous-wave (CW) excitation at 1.732 eV. The peak at 3.46 eV arises from SHG, below which upconverted PL (UPL) is observed. The bottom-left inset indicates the Auger-like process responsible for populating exciton states from the higher-lying conduction band, generating UPL. The intensity of the SHG copolarized with the laser varies as the laser polarization is rotated with respect to the crystal lattice (right inset), whereas the HX UPL does not. A set of 10 narrow peaks is resolved below 3.4 eV. The top-left inset shows the PL excitation (PLE) spectrum of the HX UPL (blue), which matches the band-edge A-exciton PL spectrum (red).

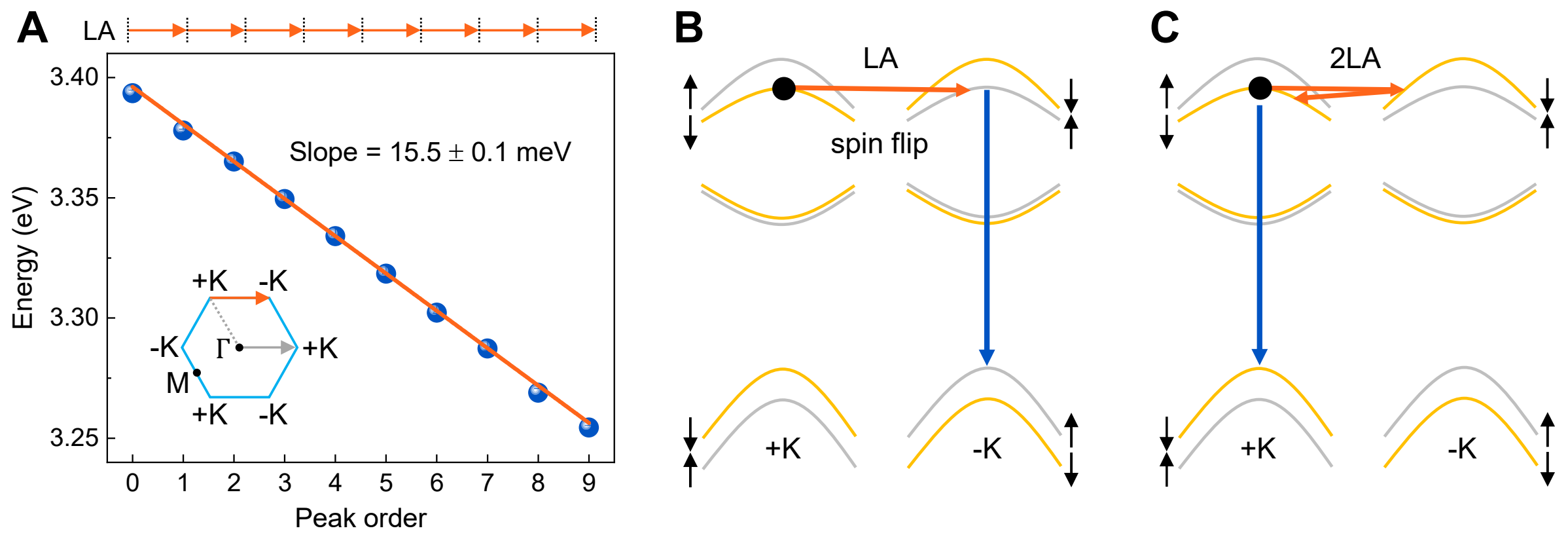

**Fig. 2. Equidistant phonon progression and phonon scattering mechanisms in high-lying exciton PL.** (A) HX UPL peak positions as a function of peak number. The slope of 15.5±0.1 meV corresponds to the calculated longitudinal acoustic (LA) phonon energy near the K-points, which allows inelastic scattering of electronic states between +K and −K points. (B), (C) The spin sublevels of conduction and valence bands of single-layer $WSe_2$ are non-degenerate at the K-points. Momentum-conserving electron-phonon scattering can involve either one phonon and a spin flip (B), or a spin-conserving two-phonon double-resonance transition (C).

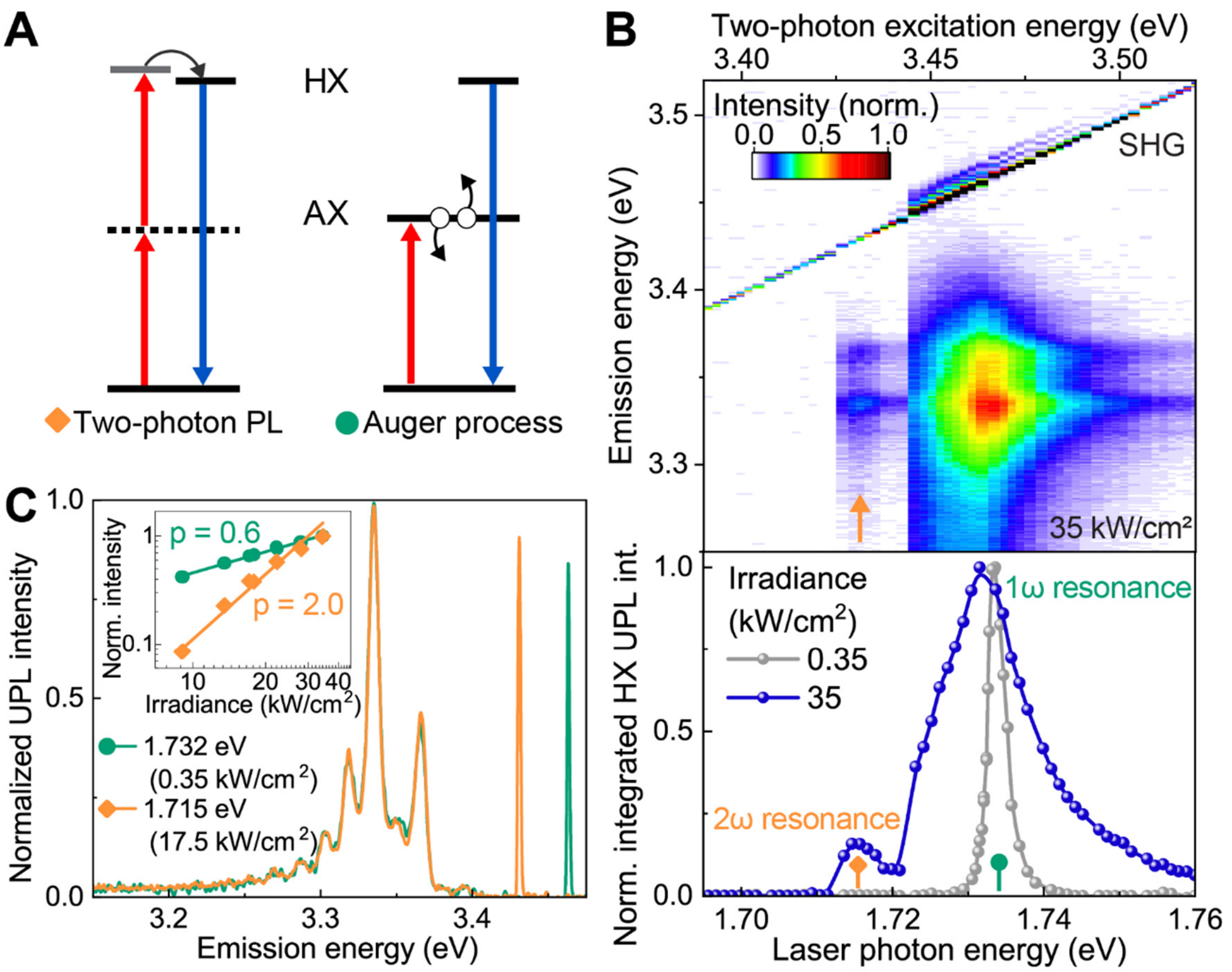

**Fig. 3. Excitation pathways of the high-lying exciton HX.** (A) HX can be formed either by two-photon absorption (TPA) involving a virtual level (dashed) or by Auger-like exciton-exciton annihilation of the A-exciton AX. TPA populates a dark *p*-like HX state, which subsequently relaxes to the bright *s*-like HX state. (B) Change of the UPL spectrum with CW excitation photon energy at an irradiance of 35 kW/cm$^2$, showing TPA (arrow) below the onset of the Auger-like double-exciton process. The diagonal line stems from SHG from the $WSe_2$ monolayer. The spectrally integrated HX UPL intensity is plotted as a function of excitation energy in the lower panel. TPA is observed at a pump intensity of 35 kW/cm$^2$, but not at 0.35 kW/cm$^2$. (C) The two excitation mechanisms are distinguished by their power dependencies. Excitation in resonance with the band-edge A-exciton at 1.732 eV gives a sublinear dependence of UPL intensity on excitation power, due to the Auger-like mechanism which bleaches the A-exciton transition. Sub-resonant excitation at 1.715 eV yields the parabolic dependence characteristic of TPA. The HX PL spectrum is independent of excitation pathway (green and orange lines).

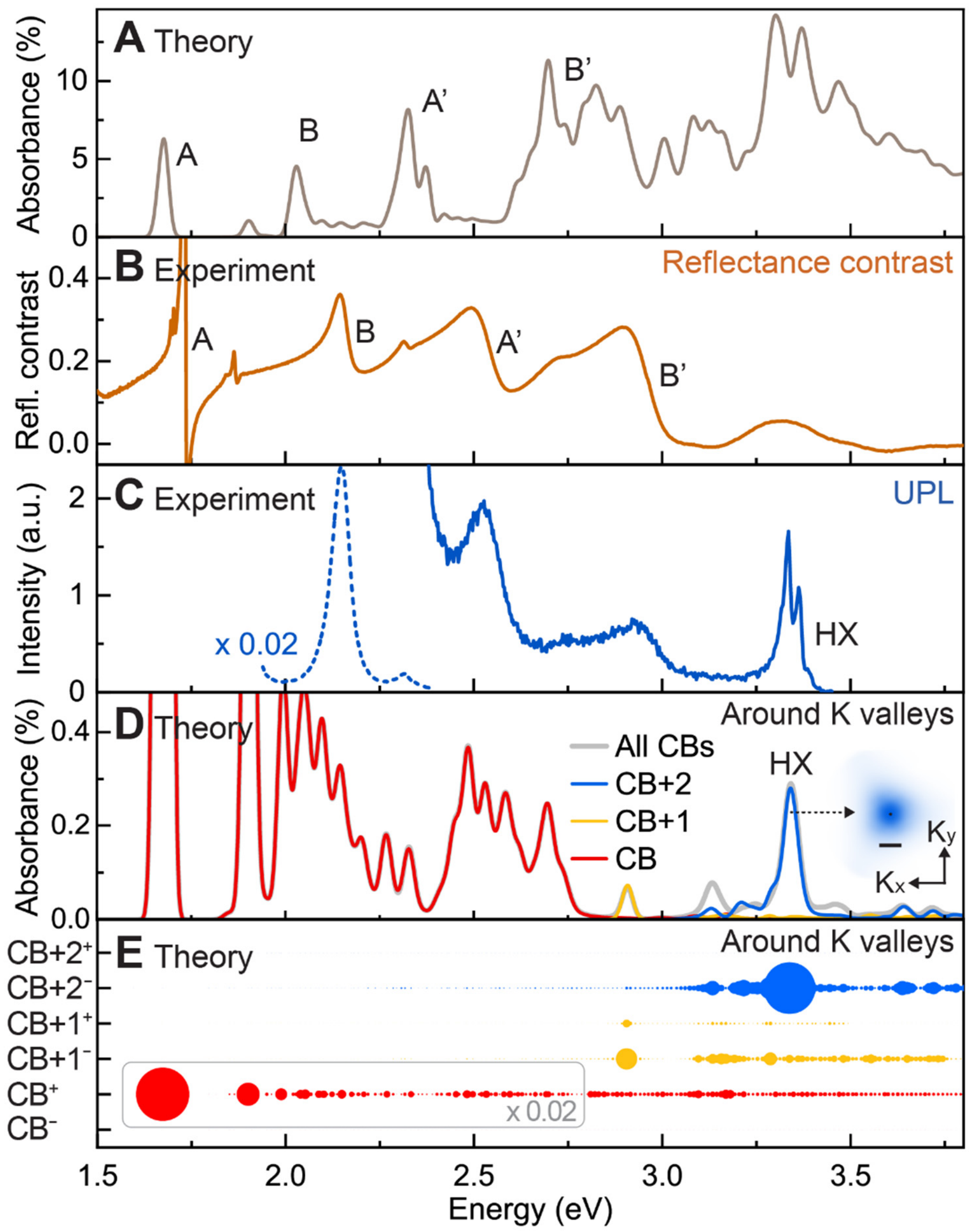

**Fig. 4. Identification of the bound HX in *ab initio GW*-BSE calculations.** (A) Calculated absorbance spectrum of bare monolayer $WSe_2$, accounting for transitions over the entire Brillouin zone. The dominant excitonic transitions A, B, A', and B' are resolved. (B) Experimental reflectance contrast of monolayer $WSe_2$ encapsulated by hBN on a sapphire substrate at 5 K. (C) Complete UPL spectrum measured under resonant pumping of the A-exciton. (D) Calculated absorbance spectrum of bare monolayer $WSe_2$, restricted to transitions around the K-points (within 6% of the Brillouin zone area) from the topmost $VB^+$ to the bottom conduction bands (CB, red), the CB+1 bands (yellow), the CB+2 bands (blue), or all bands (grey). The HX feature at 3.35 eV appears only when contributions from CB+2 are included. The inset shows the square of the HX envelope function in reciprocal space centered around K (black dot), with a scale bar of 0.2 $Å^{-1}$. (E) Contributions of the different spin-split

conduction bands to the excitonic transitions in (D), indicated by the size of the disk area. The dominant contribution to HX comes from transitions from the $VB^{+}$ to the negative-mass band $CB+2^{-}$. The disk area within the grey box is rescaled by a factor of 0.02.

# Supplementary Materials for

## Bright excitons with negative-mass electrons

Kai-Qiang Lin, Chin Shen Ong, Sebastian Bange, Paulo E. Faria Junior, Bo Peng, Jonas D. Ziegler, Jonas Zipfel, Christian Bäuml, Nicola Paradiso, Kenji Watanabe, Takashi Taniguchi, Christoph Strunk, Bartomeu Monserrat, Jaroslav Fabian, Alexey Chernikov, Diana Y. Qiu, Steven G. Louie, John M. Lupton

Correspondence to: kaiqiang.lin@ur.de, john.lupton@ur.de

**This PDF file includes:**

**Materials and Methods**

Materials

We prepared the samples following the same procedure described previously (*3, 30*). In brief, we exfoliated the monolayer $WSe_2$ and thin layers of hBN from bulk crystals ($WSe_2$, HQ Graphene; hBN, NIMS) on PDMS films (Gel-Pak, Gel-film® X4) using Nitto tape (Nitto Denko, SPV 224P). We stamp-transferred the flakes onto either $Si/SiO_2$, sapphire or diamond substrates while heating the substrate to 65°C. The hBN-encapsulated $WSe_2$ samples were annealed under high vacuum at 150°C for 5 hours.

Experimental methods

The setup is illustrated in Fig. S9. We used an objective of 0.6 numerical aperture (Olympus, LUCPLFLN) to focus the laser onto the sample and collect the signal. The sample was placed under vacuum on the cold finger of a helium-flow cryostat (Janis, ST-500). We used a grating of 600 grooves $mm^{-1}$ or 150 grooves $mm^{-1}$ to disperse the signals, and a CCD camera (Princeton Instruments, PIXIS 100) to record them. A 50:50 beam splitter was used to separate excitation and detection pathways. The photon energy-dependent instrument response in Fig. S9B was measured with a calibration light source (LS-1-CAL, Ocean Optics) placed in front of the objective. The original spectra are shown in the main text and the supporting information, without correction for the instrument response.

We measured the photoluminescence (PL) of monolayer $WSe_2$ by exciting samples with an argon-ion laser (Spectra Physics, 2045E) at 488 nm and filtering out the laser line from the signal with a 488 nm long-pass edge filter. The reflectance contrast of monolayer $WSe_2$ was measured using a broad-band Xenon lamp (EQ-99X, Energetiq). We measured the upconverted PL and SHG of monolayer $WSe_2$ by exciting samples with a tunable continuous-wave laser (Sirah, Matisse CR) and filtering out the laser line using a 680 nm short-pass filter.

*Ab initio GW* and *GW*-BSE Calculations

Density-functional theory (DFT) Kohn-Sham wavefunctions for monolayer $WSe_2$ are calculated using the Quantum ESPRESSO (*31*) package as the starting orbital energies and wavefunctions for our *GW* calculation. For the DFT calculations, we use a plane-wave basis set and norm-conserving pseudopotentials. Scalar-relativistic (SR) pseudopotentials are used for spin-unpolarized scalar-relativistic calculations, and fully relativistic (FR) pseudopotentials are used for fully relativistic noncollinear spinor calculations. The generalized gradient approximation (GGA-PBE) (*32-34*) is used for the electron exchange and correlation energy. To accurately capture the exchange contribution to the *GW* quasiparticle (QP) self-energy of the monolayer $WSe_2$, semi-core 5*s*, 5*p*, and 5*d* states are included in the pseudopotentials of W in addition to the 6*s* and 6*p* valence electrons. The plane-wave cutoff for the DFT calculation is set to 80 Ry for the plane-wave expansion of the wavefunctions. The length of the periodic supercell is set to be $L_z$ = 120 Å. The crystal structure of monolayer $WSe_2$ (*35*) has in-plane lattice constants of 3.2820 Å and an atomic-plane to atomic-plane Se-Se distance of 3.3411 Å.

The quasiparticle (QP) self-energies and exciton excitation energies are, respectively, computed using the *ab initio GW* and *GW* plus Bethe-Salpeter equation (*GW*-BSE) approaches as implemented in the BerkeleyGW (*27-28, 36*) package. In the *GW* and *GW*-BSE calculations, the Coulomb interaction beyond 60 Å in the *z*-direction (i.e., out-of-plane direction) is truncated to prevent spurious interactions between periodic images (*37*). The calculated QP band structure is plotted in Fig. S1 and Fig. 1A.

To calculate the excitation energies of the excitons, the Bethe-Salpeter equation (BSE) (*28, 36*) is solved by using the above-calculated QP energies and spinor DFT wavefunctions. The electron-hole interaction kernel of the BSE Hamiltonian is first calculated on a uniform **k**-grid of 72×72, using two valence and eight conduction bands and a dielectric matrix that is calculated on a uniform **q**-grid of 72×72, summed over 1400 bands and using a 5-Ry-plane-wave cutoff. The BSE Hamiltonian is diagonalized using interaction kernel matrix elements that are interpolated (*28, 38*) from the uniform 72×72 k-grid to a finer uniform k-grid of 120×120, using directly calculated matrix elements for **q**-points of a density equivalent to 150×150. Using the calculated exciton wavefunctions and excitation energies, the imaginary part of the dielectric function is calculated using $\mathrm{Im}[\epsilon(\omega)] = \frac{8\pi^2 e^2}{\omega^2}\frac{1}{V}\sum_S |\mathbf{e}\cdot\langle 0|\mathbf{v}|S\rangle|^2 \delta(\omega - \Omega^S)$, where $\langle 0|\mathbf{v}|S\rangle$ is the optical velocity matrix element between exciton $|S\rangle$ and ground state $|0\rangle$, $\Omega^S$ is the energy of state $|S\rangle$, $\mathbf{e}$ is the direction of the polarization of light, and $V$ is the crystal volume. The velocity matrix element of the exciton, $\langle 0|\mathbf{v}|S\rangle$, is a coherent superposition of transitions between non-interacting electron-hole pairs $\langle v\mathbf{k}|\mathbf{v}|c\mathbf{k}\rangle$, given by $\langle 0|\mathbf{v}|S\rangle = -\Omega^S \sum_{vc\mathbf{k}}^S \left(A_{vc\mathbf{k}}^S \frac{\langle v\mathbf{k}|\mathbf{v}|c\mathbf{k}\rangle}{E_{v\mathbf{k}} - E_{c\mathbf{k}}}\right)$, where $E_{n\mathbf{k}}$ and $|n\mathbf{k}\rangle$ are the Kohn-Sham eigenvalues and eigenfunctions, respectively. To obtain the absorption spectrum from a selectively excited part of momentum space, we calculate the mode-decomposed (MD) imaginary part of the dielectric function, $\mathrm{Im}[\epsilon(\omega)]_{\mathrm{MD}} = \frac{8\pi^2 e^2}{\omega^2}\frac{1}{V}\sum_S |\mathbf{e}\cdot\langle 0|\mathbf{v}|S\rangle_{\mathrm{MD}}|^2 \delta(\omega - \Omega^S)$, where $\langle 0|\mathbf{v}|S\rangle_{\mathrm{MD}} = -\Omega^S \sum_{vc\mathbf{k}\in\mathrm{subset}}^S \left(A_{vc\mathbf{k}}^S \frac{\langle v\mathbf{k}|\mathbf{v}|c\mathbf{k}\rangle}{E_{v\mathbf{k}} - E_{c\mathbf{k}}}\right)$, i.e., the band- and k-resolved contributions to the dielectric function. To calculate the absorbance, the real part of the dielectric function is firstly obtained from the imaginary part using the Kramers-Kronig relation. Using the dielectric function $\epsilon(\omega)$, we calculate the extinction coefficient $\kappa(\omega)$. Finally, the absorbance spectra A(ω), as defined by $\mathrm{A}(\omega) = \log_{10}[I(\omega)/I_0(\omega)] \times 100\%$, are plotted in Fig. 4 of the main text and Fig. S7.

In Fig. 4D, the plots are made for contributions to the absorbance due to the subset of $vc\mathbf{k}$ that corresponds to parts of momentum space that are circular patches of 0.2 $\text{Å}^{-1}$ radius centered at the K-valleys (highlighted in Fig. S1A), involving only transitions from the highest valence band. In Fig. 4E, we plot the mode-decomposed oscillator strengths for each contributing conduction band, where the area of each disk is proportional to the integrated oscillator strengths from contributing excitons, where the oscillator strength of each excitonic state, $|S\rangle$, is given by $f_S = \frac{2|\mathbf{e}\cdot\langle 0|\mathbf{v}|S\rangle_{\mathrm{MD}}|^2}{\Omega^S}$. To analyze the exciton radius, we calculate the root mean square radius of the exciton envelope function in real space. In Fig. S8C, we show the momentum-space envelope functions of three individual excitons arising primarily from CB, CB+1, and CB+2 in the vicinity of the K-valleys. Notably, the CB band curvature is smaller than the CB+1 band curvature, and CB+2 has negative curvature. Therefore, the reduced mass of the CB exciton is smaller than that of the CB+1 exciton, and the reduced mass of the CB+1 exciton is smaller than that of the CB+2 exciton (HX). This increase in mass is consistent with the fact that the CB exciton has a larger exciton radius than the CB+1 exciton and the CB+1 exciton has a larger exciton radius than the CB+2 exciton. The CB+2 exciton also has a larger binding energy of 0.6 eV than the CB+1 exciton (0.55 eV) and the CB exciton (0.45 eV). These values will decrease with hBN encapsulation because of increased dielectric screening.

Effective-mass model

In order to investigate the stability of the HX within the effective-mass limit, we restrict ourselves to the two bands with the largest contribution, i.e. $\mathrm{VB}^+$ and $\mathrm{CB+2}^-$ around the K-points, based on the *GW*-BSE calculations given in Fig. S1, Fig. 1A and Fig. 4. The effective masses are calculated from the *GW* band structure, leading to $m_{\mathrm{h}}^{*}$ = 0.3636 for $VB^{+}$ and $m_{\mathrm{e}}^{*}$ = −0.4604 for $CB{+}2^{-}$. These values fulfill the condition $|m_{\mathrm{h}}^{*}| < |m_{\mathrm{e}}^{*}|$ and thus the reduced mass of the exciton, $\mu = m_{\mathrm{e}}^{*}m_{\mathrm{h}}^{*}/(m_{\mathrm{e}}^{*} + m_{\mathrm{h}}^{*})$, remains positive. For illustration purposes, we investigate the effect of $|m_{\mathrm{h}}^{*}| > |m_{\mathrm{e}}^{*}|$ by simply arbitrarily interchanging the two values. The excitonic states are obtained by solving the effective BSE (*28*, *39-40*), given by

$$[E_{\mathrm{e}}(\vec{k}) - E_{\mathrm{h}}(\vec{k}) - \Omega_N]A_N(\vec{k}) + \sum_{\vec{k'}} V(\vec{k} - \vec{k'})\, A_N(\vec{k'}) = 0 \qquad (1),$$

where $E_{\mathrm{e}}(\vec{k}) = E_0 + \hbar^2k^2/2m_e^*$, $E_{\mathrm{h}}(\vec{k}) = -\hbar^2k^2/2m_{\mathrm{h}}^*$, and $A_N(\vec{k})$ is the envelope function of the N-th exciton state. The electron-hole interaction $V(\vec{k} - \vec{k'})$ is described by the Rytova-Keldysh potential (*41-43*)

$$V\left(\vec{k} - \vec{k}'\right) = \frac{-1}{\mathcal{A}} \frac{\mathrm{e}^2}{2\varepsilon_0} \frac{1}{\varepsilon\left|\vec{k} - \vec{k}'\right| + \mathrm{r}_0\left|\vec{k} - \vec{k}'\right|} \qquad (2),$$

in which $\mathcal{A}$ is the unit area, e is the electron charge, $\varepsilon_0$ is the vacuum permittivity, $r_0$ is the screening length of the 2D material, and ε is the effective dielectric constant. For the calculations, we assume a bare $WSe_2$ monolayer, i.e. $\varepsilon = 1$ and $\mathrm{r}_0 =$ 45.1 Å (*44*). We solve the effective BSE numerically in a 2D *k*-grid from -0.5 to 0.5 $Å^{-1}$ in $k_x$ and $k_y$ directions with 121 x 121 points (leading to a spacing between *k*-points of $\Delta k$ = 0.5/60 ≈ 0.0083 $Å^{-1}$). To improve convergence, we average the Coulomb potential around each *k*-point in a square region of $-\Delta k/2$ to $\Delta k/2$ sampled with 121 x 121 points.

In Fig. S2A, we present the electronic band structure within the effective-mass approximation for the two conditions $|m_{\mathrm{h}}^{*}| < |m_{\mathrm{e}}^{*}|$ (solid blue curves) and $|m_{\mathrm{h}}^{*}| > |m_{\mathrm{e}}^{*}|$ (dashed grey curves). In Fig. S2B we show the energy difference between conduction and valence bands, $E_{\mathrm{e}}(\vec{k}) - E_{\mathrm{h}}(\vec{k})$, i.e. the diagonal contribution of the BSE equation in Eq. (1). The two different mass conditions thus lead to distinct effective curvatures. From this analysis we can already envision localized excitons around $k = 0$ for $|m_{\mathrm{h}}^{*}| < |m_{\mathrm{e}}^{*}|$, whereas for $|m_{\mathrm{h}}^{*}| > |m_{\mathrm{e}}^{*}|$ the excitons would be delocalized at the edges of the *k*-region we considered in the calculations. The fundamental exciton ($N$ = 1) envelope function is shown in Fig. S2C, supporting our expectations of bound excitons around $k = 0$ for the condition of $|m_{\mathrm{h}}^{*}| < |m_{\mathrm{e}}^{*}|$ and no bound exciton for the condition of $|m_{\mathrm{h}}^{*}| > |m_{\mathrm{e}}^{*}|$. We note that the *s*-like shape of the exciton envelope function obtained within this effective description is consistent with the results obtained from the *ab initio GW*-BSE calculations.

Phonon calculation and electron-phonon coupling

Based on state-of-the-art DFT, we perform first-principles calculations using the Vienna *ab-initio* simulation package (VASP) (*45*). We use the projector-augmented-wave potential method with tungsten $5p^65d^46s^2$ and selenium $4s^24p^4$ valence states, together with the PBE-GGA parameterization for the exchange correlation functional (*32*). A plane-wave basis set with a kinetic energy cutoff of 500 eV and a 15×15 ***k*** mesh over the electronic Brillouin zone (BZ) leads to converged results. For structural relaxation of the unit cell, we consider energy differences converged to within $10^{-6}$ eV and Hellmann-Feynman forces converged to within $10^{-4}$ eV/Å. We obtain the harmonic interatomic force constants with density functional perturbation theory using a 5×5 supercell with a ***k***-point sampling of 3×3 of the electronic BZ. The phonon dispersion and eigenvectors are calculated using the PHONOPY package (*46*).

We estimate the strength of electron-phonon coupling by building a smallest-possible non-diagonal supercell that includes the K-point in the vibrational BZ (*47*):

$$\begin{bmatrix}\mathbf{a}_\mathrm{s}\\ \mathbf{b}_\mathrm{s}\end{bmatrix} = \begin{bmatrix}1 & 2\\ 0 & 3\end{bmatrix}\begin{bmatrix}\mathbf{a}_\mathrm{p}\\ \mathbf{b}_\mathrm{p}\end{bmatrix} \quad (3),$$

where $\mathbf{a}_\mathrm{s}$ and $\mathbf{b}_\mathrm{s}$ are the lattice parameters of the supercell, and $\mathbf{a}_\mathrm{p}$ and $\mathbf{b}_\mathrm{p}$ are the lattice parameters of the unit cell. Using this supercell, we calculate the electronic energy levels at the electronic K-points as a function of the normal mode amplitude $u_{\boldsymbol{q}v}$ of the LA phonon mode at the vibrational K-point. The normal-mode amplitude is defined as (*48*)

$$u_{\boldsymbol{q}v} = \frac{1}{\sqrt{N_p}} \sum_{\boldsymbol{R}_p,\alpha,i} \sqrt{m_\alpha} h_{p\alpha i} e^{-i\boldsymbol{q}\cdot\boldsymbol{R}_p} w_{-\boldsymbol{q}v;i\alpha} \quad (4),$$

where $\boldsymbol{q}$ and $v$ are reciprocal-space phonon wavevector and branch, respectively, $N_p$ is the number of primitive cells in the real-space supercell, $\boldsymbol{R}_p$ is the position vector of unit cell $p$, $m_\alpha$ is the nuclear mass of atom $\alpha$, $i$ runs over Cartesian coordinates, $h_{p\alpha i}$ is the displacement coordinate, and $w_{-\boldsymbol{q}v;i\alpha}$ is the corresponding eigenvector.

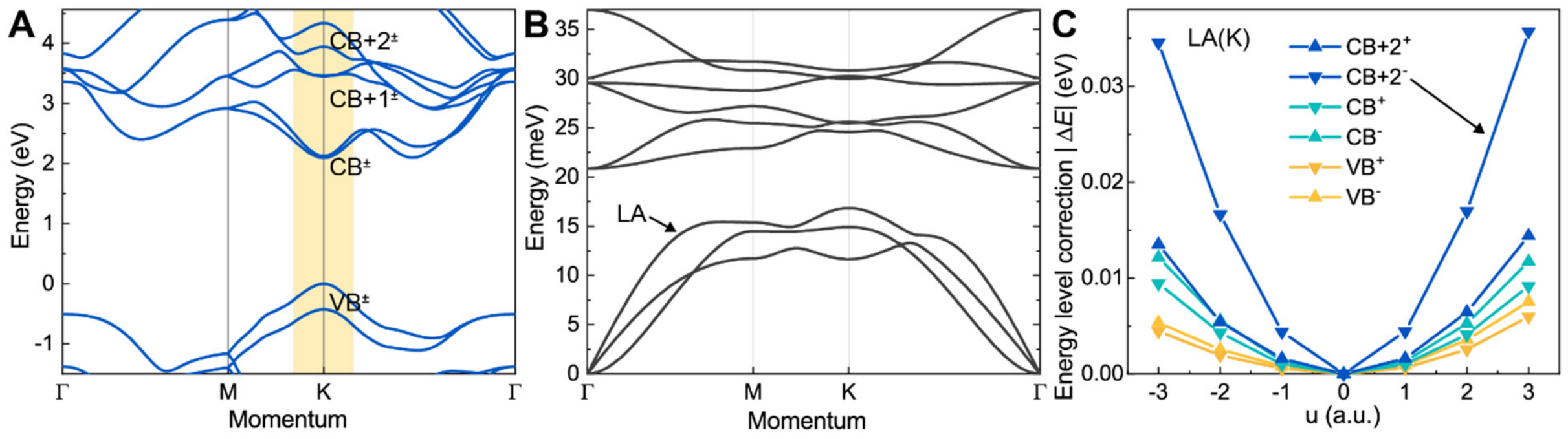

**Fig. S1.**

*GW* band structure (A) and calculated phonon dispersion (B) of free-standing monolayer $WSe_2$. The bands around the K-point are labelled with CB+2$^+$, CB+2$^-$, CB+1$^+$, CB+1$^-$, CB$^+$, CB$^-$, VB$^+$ and VB$^-$ according to their energy. The yellow shading highlights the momentum space used to identify the high-lying exciton (HX) in the absorbance spectrum in Fig. 4D using *GW*-BSE calculations. (C) Energy level correction of electronic states at the K-point with respect to the LA(K) phonon, showing much stronger electron-phonon coupling for electronic states in CB+2$^-$ than in CB$^+$ and VB$^+$. "$u$" is the normal mode amplitude of the LA(K) phonon as defined in Eq. (4).

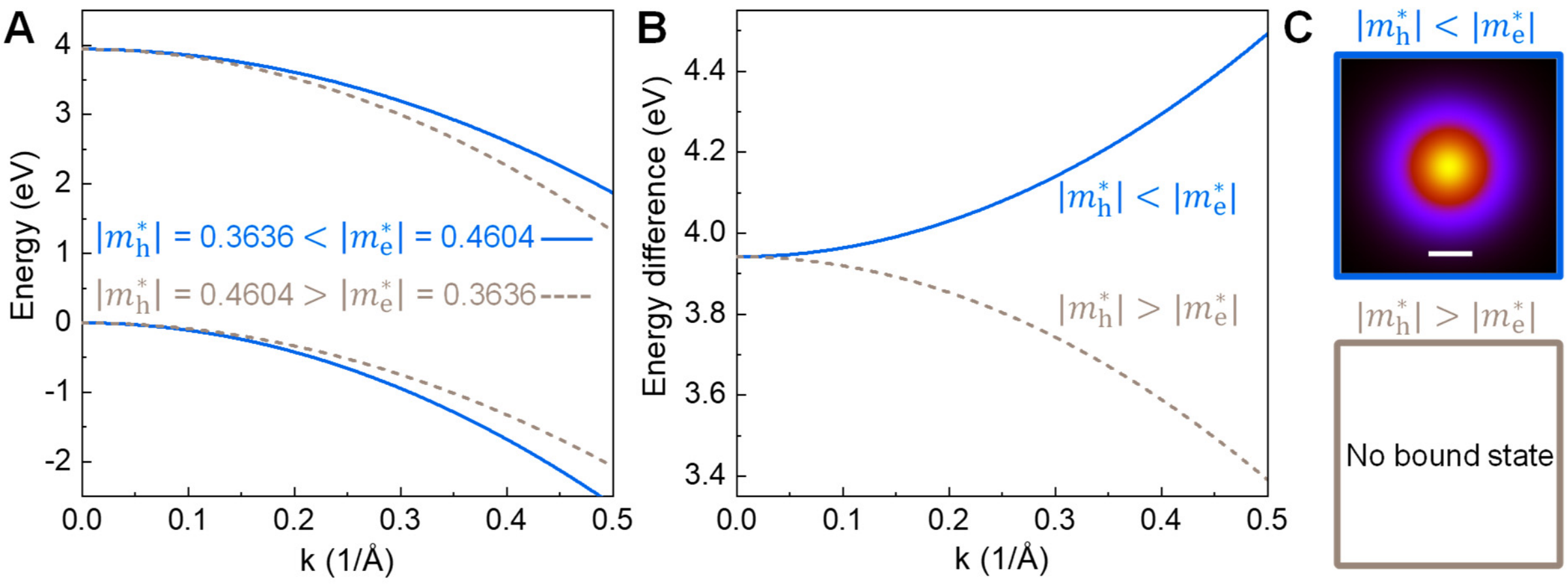

**Fig. S2.**

Hydrogen-like model of an exciton using a parabolic band dispersion for the (positive-mass hole) valence band and (negative-mass electron) conduction band. (A) The band dispersions are examined for two different conditions: $|m_h^*| < |m_e^*|$ (solid blue curves) and $|m_h^*| > |m_e^*|$ (dashed grey curves). (B) The corresponding energy difference of conduction and valence bands, $E_e(\vec{k}) - E_h(\vec{k})$, for the two conditions considered. (C) Probability density $|A(\vec{k})|^2$ of the ground-state exciton wavefunction under the two-mass conditions. The condition of $|m_h^*| < |m_e^*|$ leads to a stable bound exciton localized at $k = 0$ (i.e., in the K-valleys). In contrast, the condition of $|m_h^*| > |m_e^*|$ does not support a bound state at $k = 0$. The scale bar is 0.2 Å$^{-1}$.

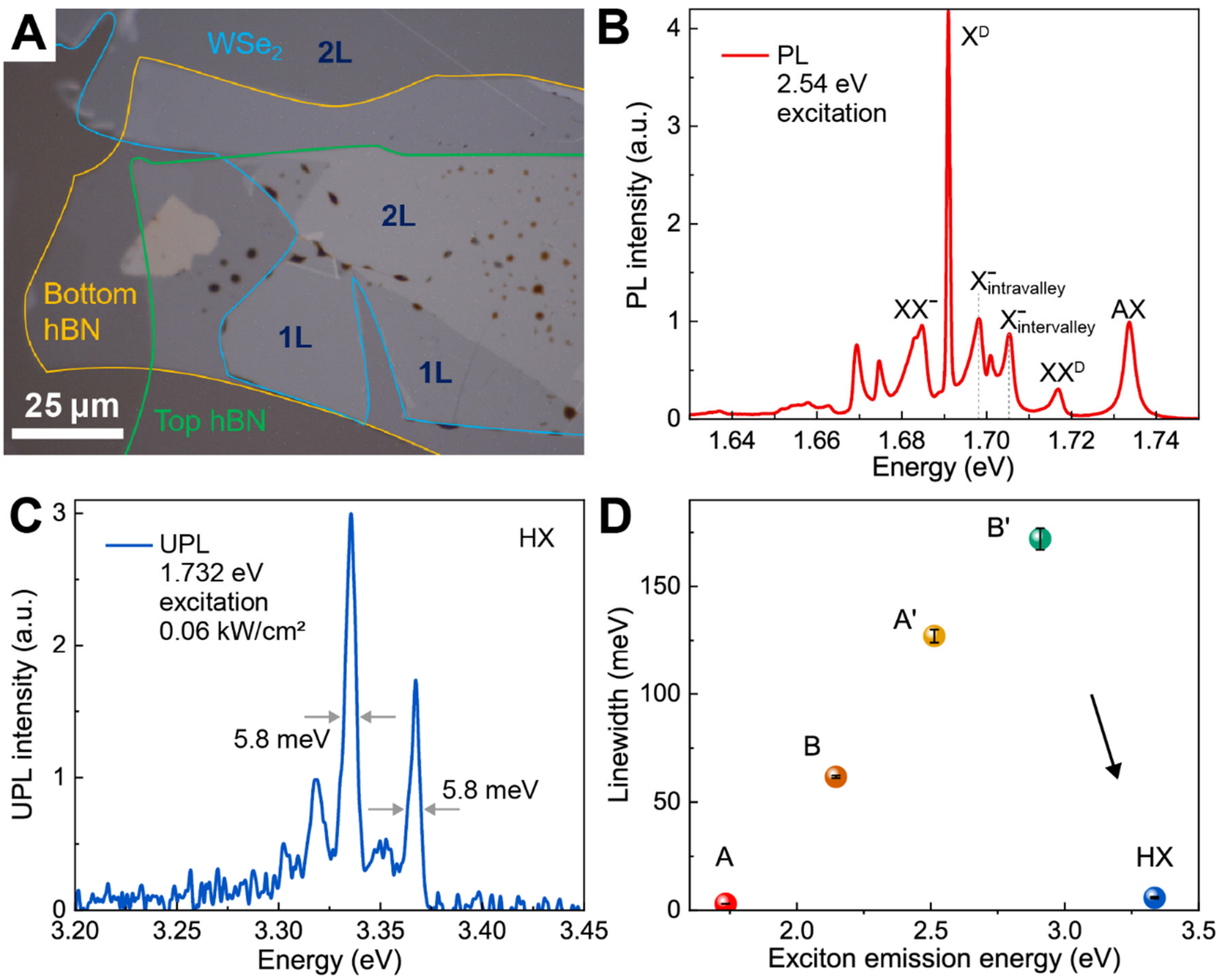

**Fig. S3.**

(A) Optical microscopic image of the hBN-encapsulated $WSe_2$ (Sample 0) on a sapphire substrate, the sample used in the experiments described in the main text. (B) PL spectrum of the hBN-encapsulated monolayer $WSe_2$ in (A) excited by a continuous wave laser at 488 nm. The A-exciton (AX), biexciton ($XX^D$), trions ($X^-_{intravalley}$, $X^-_{intervalley}$), dark exciton ($X^D$) and charged biexciton ($XX^-$) are labelled. (C) UPL of the high-lying exciton measured with a 600 grooves/mm high-resolution grating showing peaks with 5.8 meV full width at half maximum. (D) In contrast to the HX, the PL linewidth of lower-energy excitons tends to increase with transition energy. Energies and linewidths (FWHM) of the excitonic states B, A', and B' are extracted from the UPL spectrum in Fig. 4. A detailed analysis of exciton PL linewidth is provided in Table S1.

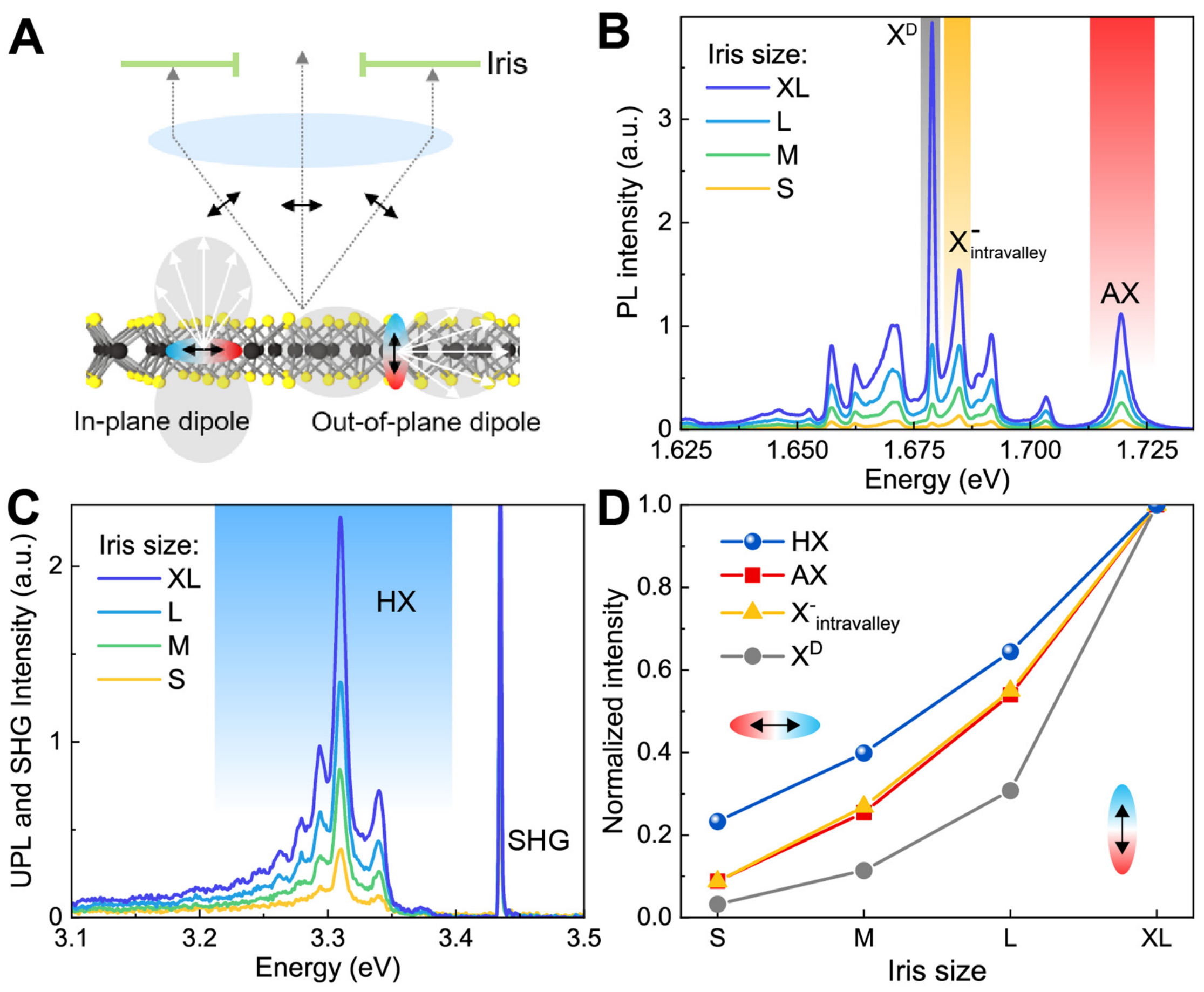

**Fig. S4.**

Distinguishing in-plane and out-of-plane dipole orientations of excitons in hBN-encapsulated monolayer $WSe_2$. Spin-allowed transitions are optically "bright" and have in-plane dipoles, while spin-forbidden transitions are nominally "dark" and only couple with light through out-of-plane transition dipoles (*49-51*). (A) Illustration of the measurement of the dependence of exciton emission on the size of an iris placed in the Fourier plane. A small iris will select in-plane dipoles whereas a larger iris will also allow light from out-of-plane dipoles to pass. (B) Iris size dependence of the PL spectrum of transitions close to the band gap. Four iris sizes were chosen and, for simplicity, are labeled qualitatively as small (S), medium (M), large (L), and extra-large (XL). (C) Dependence of the HX UPL spectrum on iris size. (D) Dependence on iris size of the spectrally integrated excitonic emission (marked red, yellow, grey in (B) and blue in (C)). The dependence is much stronger for the dark exciton $X^D$ than for the neutral A-exciton AX and the trion $X^-_{intravalley}$, since $X^D$ is characterized by an out-of-plane dipole. The dependence on iris size is much weaker for HX than for $X^D$, implying that HX emits via an in-plane dipole.

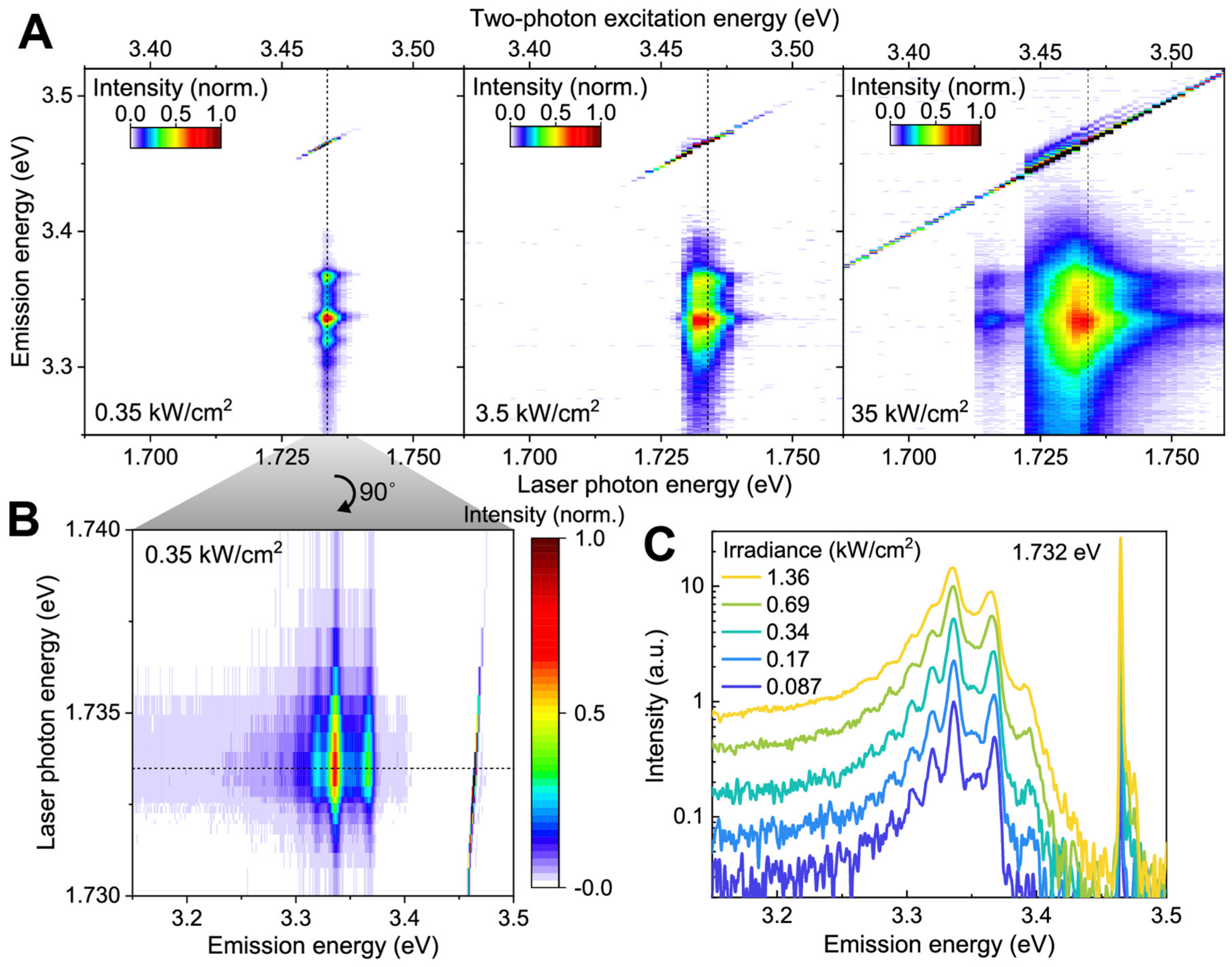

**Fig. S5.**

(A) Dependence of the UPL PLE spectrum on pump intensity for irradiances of 0.35 kW/cm$^2$, 3.5 kW/cm$^2$ and 35 kW/cm$^2$. (B) Rescaled and rotated close-up of the UPL PLE spectrum at a pump irradiance of 0.35 kW/cm$^2$. (C) Power dependence of UPL spectra at an excitation photon energy of 1.732 eV. No significant change of the spectral shape of HX emission with excitation intensity is observed.

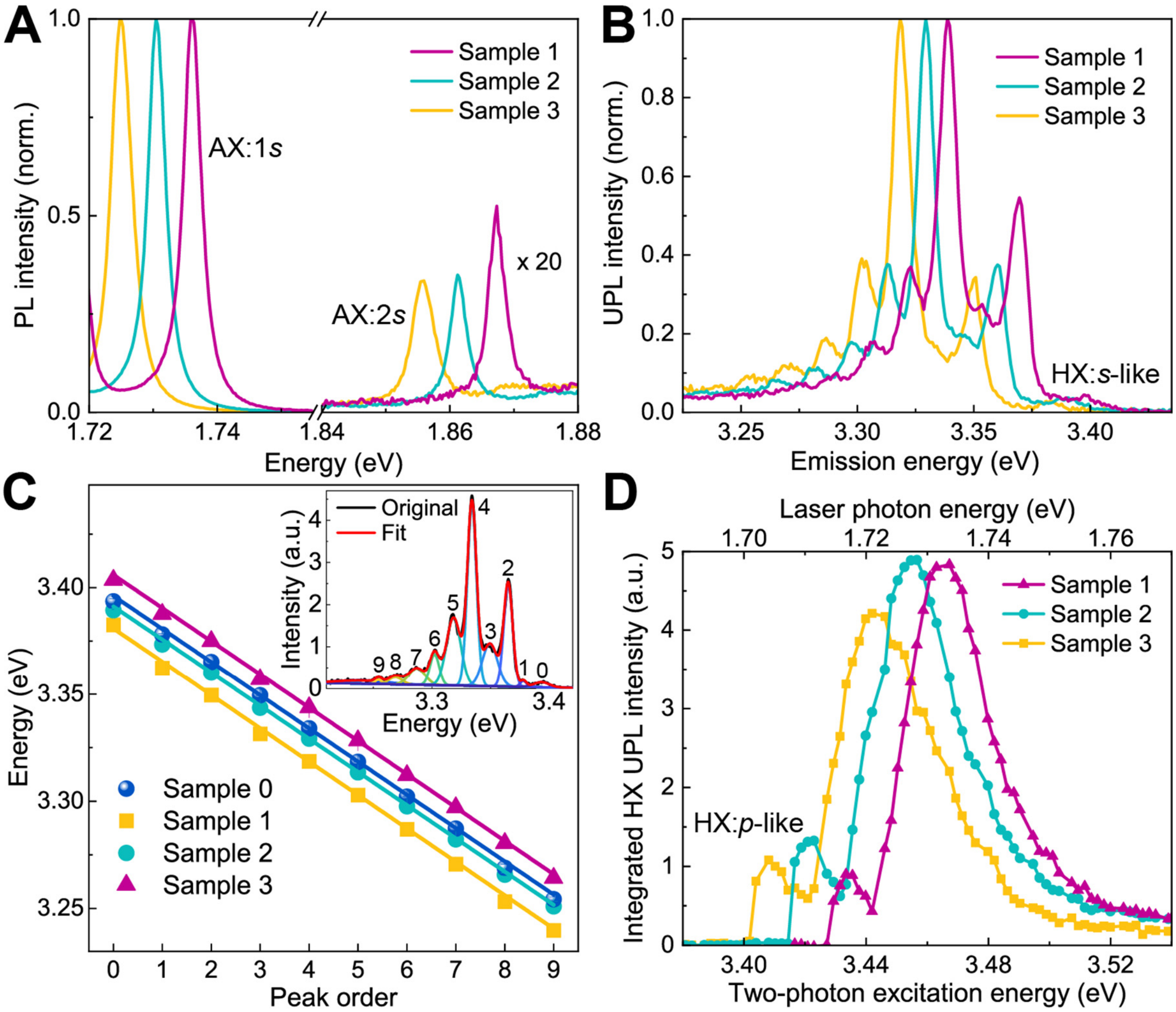

**Fig. S6.**

Correlation between the A-exciton (AX) PL and high-lying exciton (HX) PL for different dielectric environments of the $WSe_2$ monolayer. Samples 1, 2 and 3 correspond to three different $WSe_2$ monolayers encapsulated in hBN with different thicknesses. (A) PL of 1*s* and 2*s* states of the A-exciton. (B) UPL of the high-lying exciton with the lowest-order peak (i.e. peak 0 in Fig. 2) labelled as the *s*-like HX state. (C) HX UPL peak positions as a function of peak order obtained through a ten-Gaussian global fit to the three spectra of panel (B). The inset shows the fit of Sample 0 in the main text and the result (blue spheres, the same as Fig. 2A) is shown for reference. Linear fits of peak position against peak order determine the spacing of the phonon progression to be 15.50±0.10 meV (Sample 0), 15.59±0.12 meV (Sample 1), 15.50±0.07 meV (Sample 2) and 15.56±0.07 meV (Sample 3), as listed in Table S2. (D) Excitation spectra of the UPL and the two-photon PL at the same irradiance of 35 kW/cm$^2$, showing the energy of the *p*-like HX state. The lower abscissa indicates double the excitation photon energy. Details of the peak positions are summarized in Table S2.

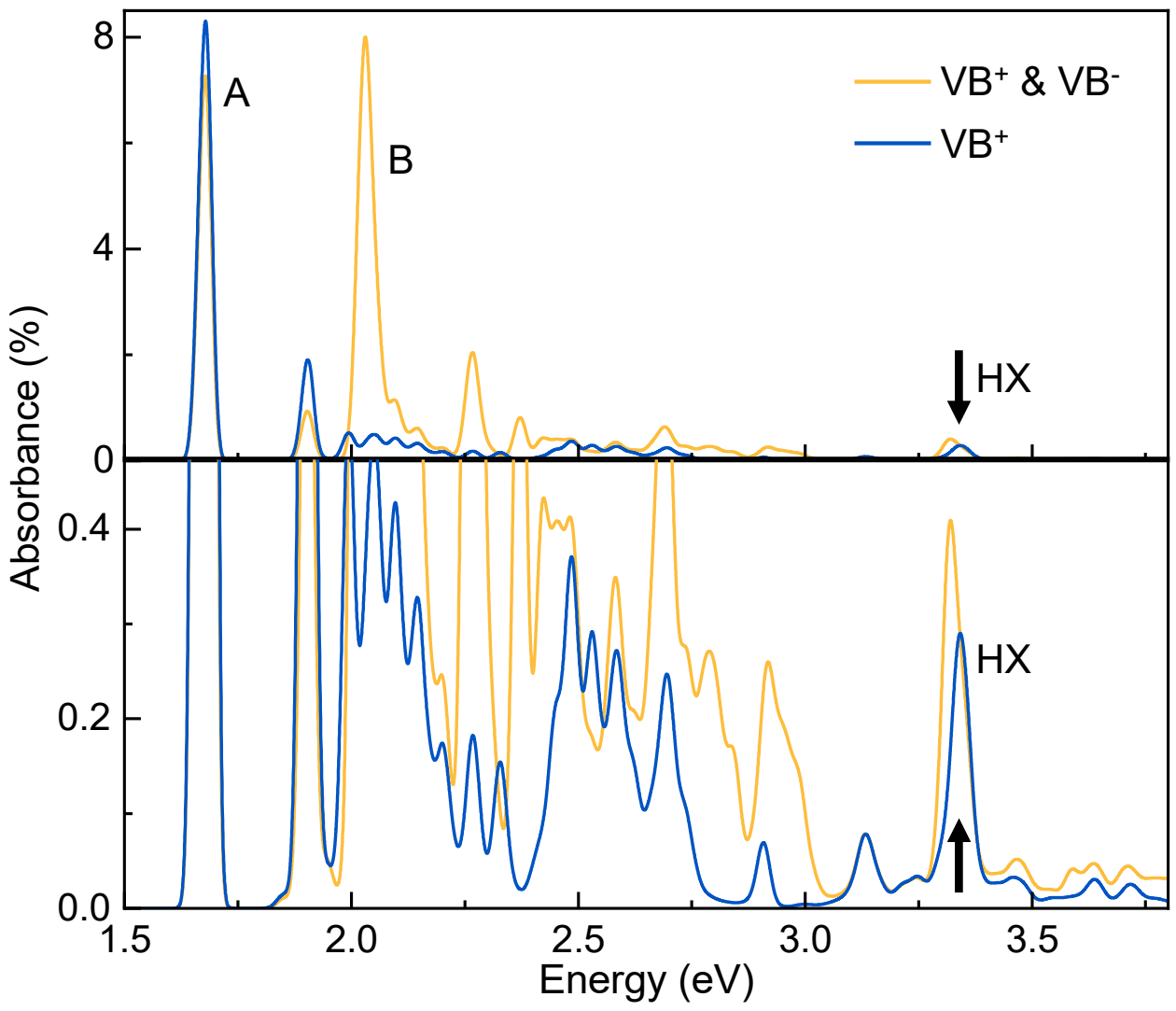

**Fig. S7.**

Contributions of the spin-split valence band to exciton formation. Calculated absorbance spectrum (in arbitrary units) of bare monolayer $WSe_2$ from transitions around the K-points from either $VB^+$ (blue) or $VB^{\pm}$ (yellow) to all eight conduction bands $CB^{\pm}$, $CB+1^{\pm}$, $CB+2^{\pm}$ and $CB+3^{\pm}$. Top panel: plot showing full absorbance spectrum. Bottom panel: close up to show more clearly the absorbance near the HX transition. In addition to the HX from $VB^+$, a shoulder from $VB^-$ with much weaker oscillator strength overlaps with the HX (black arrow).

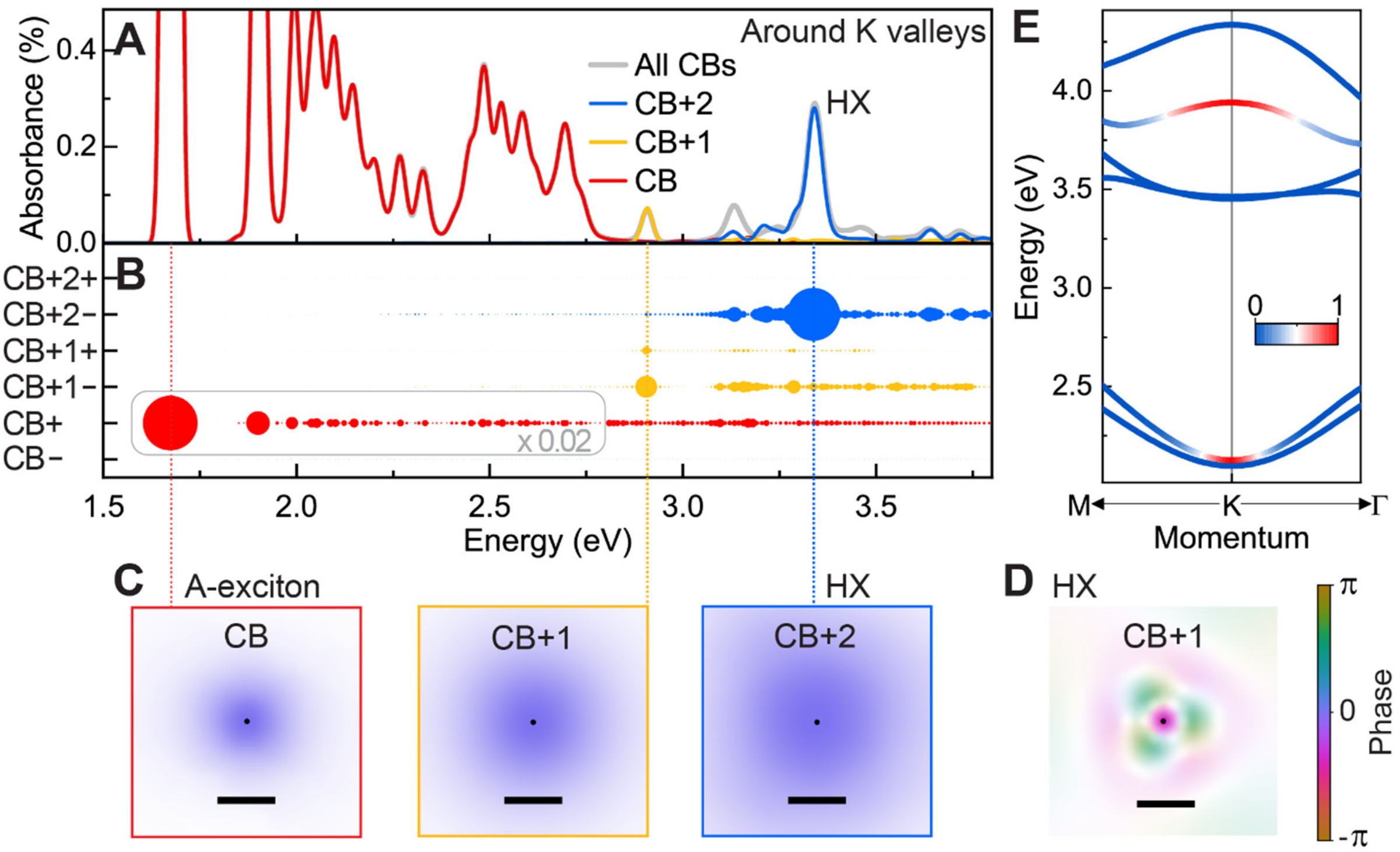

**Fig. S8.**

Exciton wavefunction envelopes. (A) Calculated absorbance spectrum reproduced from Fig. 4D. (B) Band contributions to the absorbance spectrum, reproduced from Fig. 4E. (C) Momentum-space envelope functions of three excitons with dominant contributions from CB (A-exciton), CB+1 and CB+2 (HX). The opacity is proportional to the amplitude of the envelope function and the color denotes the phase according to the scale bar. The black dot marks the K-point. The HX evidently has a larger radius in reciprocal space and thus a smaller Bohr radius compared with the A-exciton. The scale bar is 0.2 $Å^{-1}$. (D) Even though the HX transition has a dominant contribution from CB+2, the HX also has certain wavefunction amplitudes from CB+1. The figure shows the momentum-space envelop function of the CB+1 contribution to the HX. The wavefunction shows nodes revealing its *p*-like (i.e. odd-parity) non-emissive nature. The contribution of CB+1 to the HX in the absorbance spectrum is therefore minor as seen from the yellow curve in panel B. (E) Conduction bands of monolayer $WSe_2$ with the projection of the modulus-squared exciton envelope function of HX onto the CB+2$^-$ band and of AX onto the CB$^-$ band, normalized at the K-point. The momentum space is set to the range used to identify the HX in the calculated absorbance spectrum in (A), i.e. within a range of 0.2 $Å^{-1}$ around the K-points corresponding overall to 6 % of the total area of the Brillouin zone. The HX is clearly localized around the K-points, as seen by the red amplitudes in valence and conduction bands. The effective-mass approximation is therefore still valid.

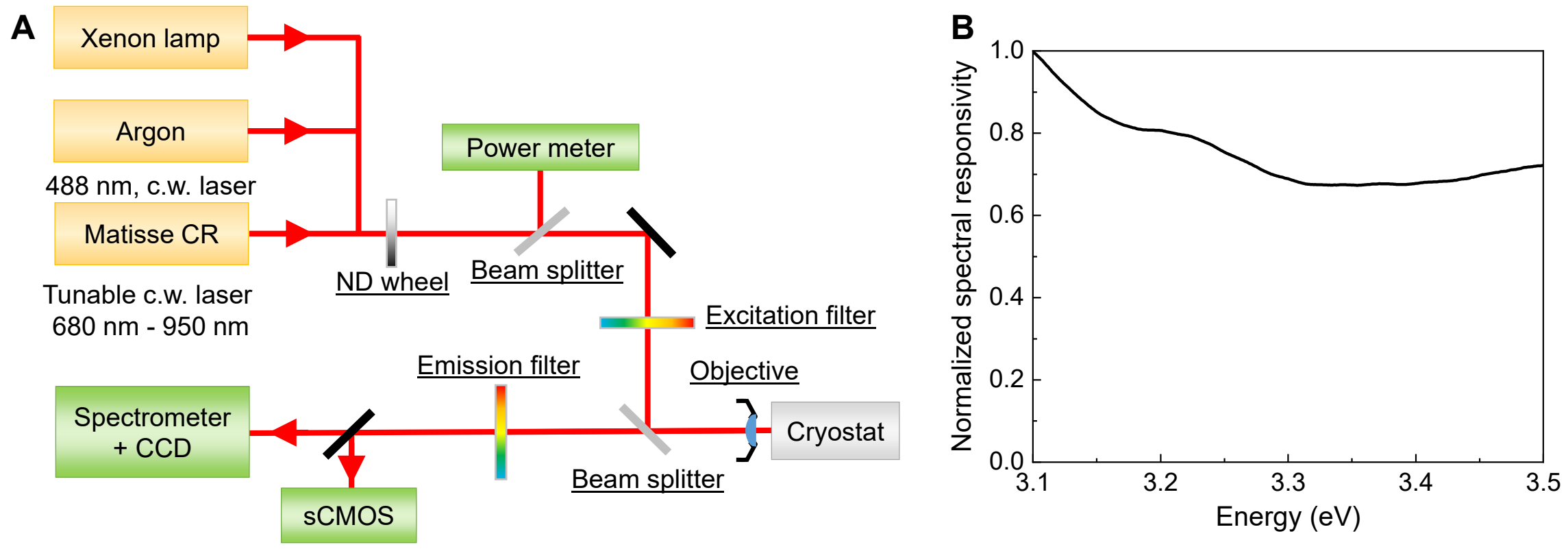

**Fig. S9.**

(A) Illustration of the experimental setup. (B) Experimentally determined spectral responsivity of the instrument.

**Table S1.**

Energies and linewidths (FWHM) of the excitonic states A, B, A’, B’ and HX extracted from the PL and UPL spectra in Fig. 4 and Fig. S3.

| Excitons | A | B | A’ | B’ | HX |
|---|---|---|---|---|---|
| Energy (eV) | 1.734 | 2.146 | 2.513 | 2.908 | > 3.2 |
| Linewidth (meV) | 3.02 ± 0.02 | 61.7 ± 0.6 | 127 ± 3 | 172 ± 5 | 5.8 ± 0.4 |

**Table S2.**

Energies of the excitonic states AX, HX:*s*-like (zero-phonon line) and HX:*p*-like, the energy difference between these states $\Delta E_{HX:s\text{-}p}$, and the phonon progression spacing $\Delta E_{phonon}$ from three different samples, in units of eV.

| Sample | AX | HX:*s*-like | HX:*p*-like | $\Delta E_{HX:s\text{-}p}$ | $\Delta E_{phonon}$ |
|---|---|---|---|---|---|
| 1 | 1.7361 | 3.3980 | 3.4350 | 0.0370 | 0.01559±0.00012 |
| 2 | 1.7305 | 3.3894 | 3.4216 | 0.0322 | 0.01550±0.00007 |
| 3 | 1.7250 | 3.3818 | 3.4091 | 0.0273 | 0.01556±0.00007 |